\documentclass{iau} 
\usepackage{graphicx}
\usepackage{natbib,xcolor}

\pubyear{2022}
\volume{372}
\jname{The Era of Multi-Messenger Solar Physics}
\editors{G.~Cauzzi \& A.~Tritschler, eds.}

\title[How the solar wind evolved to become what it is today]{How has the solar wind evolved to become what it is today?}

\author[Vidotto]{A.~A.~Vidotto}

\affiliation{Leiden Observatory, Leiden University, PO Box 9513, 2300 RA Leiden, The Netherlands }

\begin{document}

\maketitle

\begin{abstract}
In this contribution, I briefly review the long-term evolution of the solar wind (its mass-loss rate), including the evolution of observed properties that are intimately linked to the solar wind (rotation, magnetism and activity). I also briefly discuss implications of the evolution of the solar wind on the evolving Earth. I argue that studying exoplanetary systems could open up new avenues for progress to be made in our understanding of the evolution of the solar wind. 
\end{abstract}

              
\section{Introduction}
The Sun is the best studied star in the whole Universe: we can measure its properties with accuracy like no other star. However, all this information just tells us about how the Sun looks like now. Understanding how the winds of cool stars evolve is key to understand, for example, how has the solar wind and the solar system planets evolved during the past 4 billion years or what  the implication of a strong wind is for young exoplanets. 

To understand the past, and future, evolution of the Sun, including its wind, magnetism, activity, rotation, and irradiation, we rely on data of solar-like stars, in an effort to better place the Sun and the solar wind in a stellar context. Solar-like stars form a broad group of stars with spectral types from early M  to late F  ($0.4 \lesssim M/M_\odot \lesssim 1.3$) in the main sequence. These stars have in common a convective outer envelope, which is linked to the generation of magnetism, which ultimately drives their high-energy radiation and winds. In this contribution, I review the evolution of some key ingredients associated to stellar winds.

\section{Evolution of mass-loss rates}
\begin{figure}[t]
\begin{center}
 \includegraphics[width=0.44\textwidth]{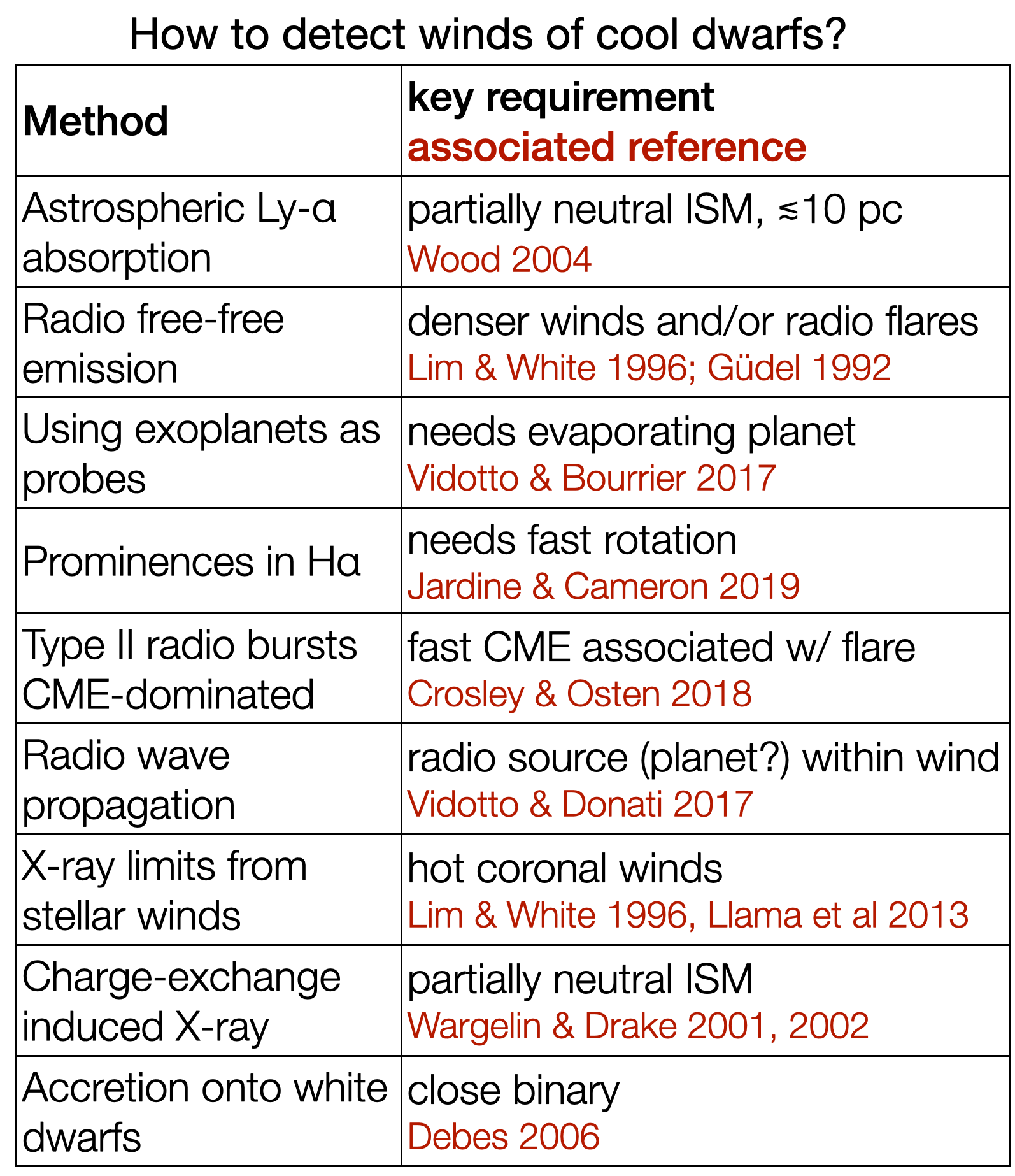} 
 \caption{Table summarising  (most of) the proposed methods to detect tenuous winds of cool dwarf stars  (adapted from \citealt{2021LRSP...18....3V}).} \label{fig1}
\end{center}
\end{figure}

\begin{figure}[t]
\begin{center}
 \includegraphics[width=0.55\textwidth]{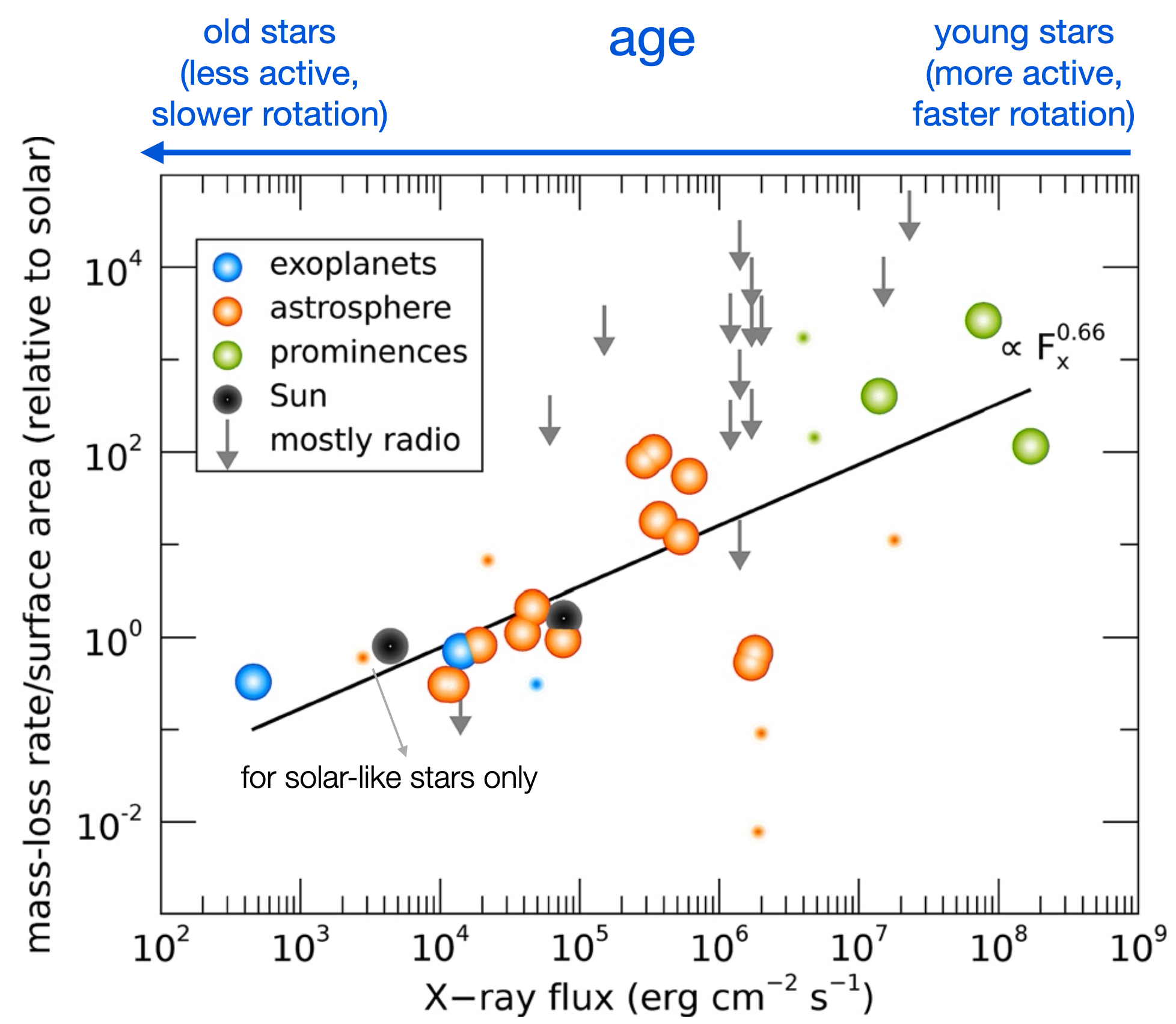} 
 \caption{Relation between  mass-loss rates and X-ray surface flux. Colour indicates the method used in the derivation of mass-loss rates (see panel on the left), with grey arrows indicating upper limits. The solid line is a power-law fit through the {larger} circles (Equation \ref{eq.mdot.fit}).  The smaller symbols are either evolved stars or M dwarfs, which were not included in the fit and neither were the stars for which only upper limits exist (arrows). Figure from \citet{2021LRSP...18....3V}.} \label{fig2}
\end{center}
\end{figure}

Detecting winds of cool dwarf stars can be very tricky, as they are in general tenuous and more difficult to detect. By modelling the observed interaction between stellar winds and the interstellar medium (ISM) \citet{2002ApJ...574..412W} proposed a correlation between X-ray flux and mass-loss rate. He showed that stars that have higher X-ray fluxes in general have winds with higher mass-loss rates. In addition to the method proposed by \citet{2002ApJ...574..412W}, other  methods have also been proposed to detect winds of cool dwarfs.  In \citet{2021LRSP...18....3V}, I reviewed some of the methods proposed so far -- this is now summarised  in Figure \ref{fig1}, while Figure \ref{fig2} \footnote{The values used in this plot were compiled from the following works: \citet{1993ApJ...406..247D, 1996ApJ...460..976L, 2000GeoRL..27..501G, 2001ApJ...547L..49W,2002ApJ...574..412W, 2005ApJ...628L.143W, 2014ApJ...781L..33W, 2010ApJ...717.1279W, 2018JPhCS1100a2028W, 2002ApJ...578..503W,  2013A&A...551A..63B, 2014Sci...346..981K,  2017A&A...599A.127F, 2017MNRAS.470.4026V,  2017A&A...602A..39V, 2019MNRAS.482.2853J, 2019ApJ...885L..30F, 2021MNRAS.500.3438O}.} shows a compilation of these results. Following the suggestion from \citet{2002ApJ...574..412W}, I present a fit for the solar-like stars (black line in Fig \ref{fig2}):
\begin{equation}\label{eq.mdot.fit}
\frac{\dot{M}}{R_\star^2} =10^{-2.75\pm 0.68} \left[ \frac{F_X}{{\rm erg\, cm}^{-2}{\rm s}^{-1}}\right]^{0.66\pm0.12}  \,\,  \frac{2 \times 10^{-14} M_\odot/{\rm yr}}{~R_{\odot}^{2}}  \, .
\end{equation}

In general, stars that have higher X-ray surface fluxes are younger (see Section 4). With age, stellar activity decreases. Therefore, the $x$-axis in Fig \ref{fig2} can also be interpreted as a roughly age indicator, whereby younger stars (which are more active and with faster rotation) are shown to the right while the older stars (which are less active and show slower rotation) are shown to the left of the plot. Using the relation from  \citet{2007LRSP....4....3G}, who found that  $L_X \propto t^{-1.5\pm 0.3}$ (for solar-like stars in the non-saturated regime) and Eq.~(\ref{eq.mdot.fit}), we have that 
\begin{equation} 
\dot{M} \propto {\rm age}^{-1.5 \times 0.66} \sim 1/{\rm age} \,\,\,\,  ({\rm roughly})\, . %
\end{equation}
Therefore, the mass-loss rates of solar-like stars, in the non-saturated regime, should decrease with age. 

\section{Stellar magnetism evolution}
Because of winds of cool dwarf stars are magnetic in origin (similar to the solar wind), magnetic field measurements are one of the essential input quantities in models of stellar winds\footnote{What makes solar-like stellar winds so effective at removing angular momentum is the stellar magnetic field \citep{1967ApJ...148..217W}.}.  In particular, it is interesting to note that the (large-scale) topology of the magnetic field evolves during solar cycle \citep[e.g.,][]{2012ApJ...757...96D,2018MNRAS.480..477V}, which affects the observed structure of the solar wind. For example, at minimum, when the solar large-scale field can be characterised by an aligned dipole, the solar wind velocity shows a bimodal structure, with fast streams coming from the polar regions, where the solar magnetic field has open field lines \citep{2008GeoRL..3518103M}. 

To measure surface magnetic fields, I highlight two different methods. (1) Zeeman-Doppler imaging (ZDI) maps the large scale field of the stellar surface, through an image reconstruction algorithm which uses as input time-series of spectropolarimetric data. (2) By measuring the broadening of magnetically sensitive lines, the Zeeman broadening technique derives the total field strength at the stellar surface. 

While the former technique is able to reconstruct the topology of the magnetic field, it can only do so on the large-scale component. While the latter technique is able to derive the total field strength (including the ZDI-undetected small-scale component), it only obtains the unsigned field (i.e., no topology indicator). As a result the two techniques are highly complementary. In depth reviews of these techniques can be found in \citet{2009ARA&A..47..333D} and \citet{2012LRSP....9....1R}. 

\begin{figure}[h]
\begin{center}
 \includegraphics[width=0.74\textwidth]{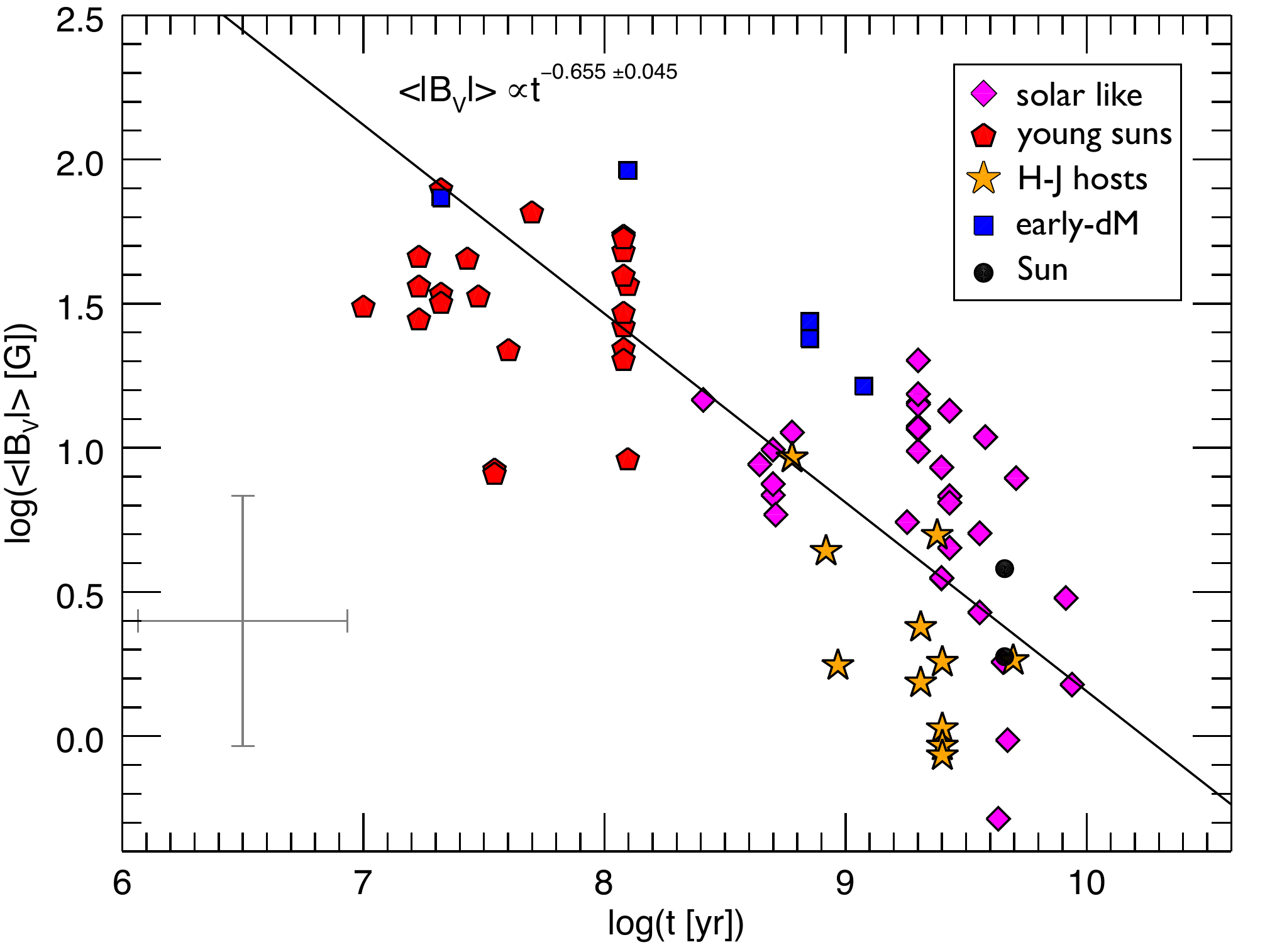} 
 \caption{Evolution of the average large scale magnetic field of a sample of solar-like stars, as measured from the ZDI technique. The large scatter shown around the fit (black line) are attributed to variability of the stellar magnetic field (e.g., cycles) as well as a large uncertainty in determining stellar ages. Figure from \citet{2014MNRAS.441.2361V}.} \label{fig3}
\end{center}
\end{figure}

While stellar magnetism has been expected to decrease with age, it was only by grouping a relatively large sample of magnetic field observations that we finally were able to quantitatively assess this decay \citep{2014MNRAS.441.2361V}. Figure \ref{fig3} shows that the large scale field of solar-like stars decay with $\sim {\rm age}^{-0.6}$, which is a rather similar slope as the ``Skumanich-law'', whereby rotation is observed to decay with $\sim {\rm age}^{-0.5}$ \citep{1972ApJ...171..565S}. Rotational evolution will be discussed next.

\section{Evolution of stellar rotation and high-energy radiation}

A key concept of winds of cool dwarf stars is that they carry away angular momentum. Therefore, during the main sequence, as a star ages, its rotational velocity decreases (spin down). On the other hand, rotation is observed to be linked with (magnetic) activity, therefore, as a star spins down, its activity also decreases. Hence, older stars are characterised by slower rotation and less intense activity. There is, thus, a feedback-loop between age, rotation and (magnetic) activity, which is mediated by the angular momentum carried away by stellar winds.

Activity is a broad term that is usually related to a number of processes. Activity can be identified in the form of spots (which can be probed through photometric monitoring). Activity can also be identified through spectroscopic lines formed in the chromosphere (e.g. CaII H\&K). Alternatively, it can be linked to coronal diagnostic measurements, such as X-ray emission and high-energy radiation in general. Note that all these processes  (spots, chromospheric and coronal emission) are magnetic in nature, albeit they are indirect measurements of magnetism. More directly, activity can be probed by measuring stellar magnetic fields (see previous section). As activity (in the broad term) decreases with age, each of these processes are also observed to decrease with age (albeit with different slopes).

In terms of rotation, there are several observational studies dedicated to mapping the evolution of surface rotation. To reduce the uncertainty in age measurements, stars in open clusters (which are coeval) have been frequently used to to measure how stellar rotation varies as a function of age \citep[e.g.][]{2009IAUS..258..363I}. In  recent years, thousands of measurements of rotation periods of low-mass stars have become available thanks to new telescopes, such as Kepler \citep[e.g.][]{2014ApJS..211...24M}. In the main sequence, stellar winds brake the rotation of stars. Therefore, models that describe  stellar rotational evolution usually parameterise the evolution of angular momentum loss through stellar winds, albeit in different scales of complexity. Examples of such models are: \citet{1988ApJ...333..236K, 2011MNRAS.416..447S, 2013A&A...556A..36G, 2015A&A...577A..28J, 2015ApJ...799L..23M, 2019A&A...631A..77A, 2021A&A...654L...5P}. 

Because high-energy radiation is linked to rotation, and high-energy radiation (in particular in the ultra-violet) is very hard to measure, in recent years, semi-empirical frameworks to model and predict the evolution of high-energy radiation have been developed \citep{2015A&A...577L...3T, 2021A&A...649A..96J, 2021A&A...654L...5P}. In general terms, these frameworks rely on models of stellar rotational evolution which are coupled to empirical relations between X-ray emission and rotation \citep[e.g.][]{2011ApJ...743...48W, 2014ApJ...794..144R} and to relations between UV radiation and X-ray emission \citep[e.g.][]{2011A&A...532A...6S,2021A&A...649A..96J}. These semi-empirical frameworks provide a much needed input for models of (exo)planetary evolution, which I will discuss next. 

\section{Concluding remarks: effects of the evolving particle, magnetic and radiation environments of solar-like stars on (exo)planets}
Understanding how the winds of cool stars evolves is key to understand wind effects on orbiting planets. A few open questions and related studies are listed below:  

\begin{itemize}
\item {\it How have the solar system planets evolved during the past 4 billion years? }There are indications that Mars lost its atmosphere because of atmospheric erosion due to a stronger wind of the young sun \citep{2007SSRv..129..207K}. Earth, on the other hand, is believed to have retained its atmosphere because of its strong intrinsic magnetic field \citep{2018MNRAS.481.5146B}, in spite of the fact that its magnetosphere would have been smaller at earlier ages \citep{2019MNRAS.489.5784C}. 
\item {\it What is the implication of a strong wind on atmospheric evaporation of close-in exoplanets?} The wind environment surrounding close-in exoplanets are harsher than the environment around planets orbiting far from their host stars \citep{2015MNRAS.449.4117V}. For unmagnetised planets, this could enhance their atmospheric evaporation \citep[][but see also \citealt{2020MNRAS.494.2417V, 2021MNRAS.500.3382C}]{2019ApJ...873...89M,2022MNRAS.509.5858H}. The interaction with a stellar wind can also shape the atmospheric material escaping from the planet \citep{2013A&A...557A.124B, 2019MNRAS.483.1481D, 2021MNRAS.501.4383V, 2022MNRAS.510.2111K}. Even if the effect of winds on erosion is not significant, it can substantially affect their observational signatures of atmospheric evaporation through spectroscopic transits \citep{2020MNRAS.498L..53C, 2021MNRAS.500.3382C}.
\item {\it How strong are the interactions between a close-in planet and the wind of its planet-hosting star?} There are different types of star-planet interactions \citep{2020IAUS..354..259V}, including interactions that are mediated by the stellar wind, such as planetary auroral radio emission \citep{2019MNRAS.485.4529K, 2021A&A...645A..59T} or planet-induced emission on the host star \citep{2020NatAs...4..577V, 2021MNRAS.504.1511K}. The strength of these interactions depend on the local (i.e., at the planet's orbit) conditions of the stellar wind, which are higher for planets orbiting closer to their host star. For this reason, star-planet interactions mediated by the wind particles and its magnetic fields are stronger in closer-in planets \citep{2016ApJ...833..140S}.
\item {\it Are stellar winds big ``cosmic ray filters''?} Cosmic rays are ionised, high-energy particles that permeate the Galaxy (in which case they are known as galactic cosmic rays) or particles that originate in stellar flares or coronal mass ejections (in which case they are know as SEPs or stellar energetic particles). Once these particles are inside the stellar wind, the magnetised  wind can modulate them both in momentum and spatial domains (what I refer above as giant cosmic ray filters). In essence, the wind can push cosmic rays out, as they are spatially advected, or can allow them in, as they diffuse into the system \citep[e.g., ][]{2013LRSP...10....3P}. These processes depend on the properties of stellar winds, which evolve in time \citep{2012ApJ...760...85C,2020MNRAS.499.2124R,2021MNRAS.504.1519R, 2021MNRAS.508.4696R}. Knowing whether cosmic rays have been more or less suppressed in the past\footnote{Also around other planetary systems \citep[e.g.,][]{2019ApJ...874...21F, 2020ApJ...897L..27H,2021MNRAS.505.1817M, 2022MNRAS.509.2091M, 2022MNRAS.515.1218M}.} are important for chemical modelling of planetary atmospheres: cosmic rays can penetrate atmospheres, providing an extra source of atmospheric ionisation, driving the formation of prebiotic molecules, and dissociating molecules \citep[e.g.][]{2013ApJ...774..108R, 2017MNRAS.465L..34A, 2021MNRAS.502.6201B}. 
\end{itemize}

The strength of some star-planet interactions (e.g., planetary auroral emission, the interplay between an evaporating atmosphere and the stellar wind, cosmic ray suppression) depends directly on the local and/or global characteristics of the stellar wind. Therefore, an interesting ``by-product'' of studying star-planet interactions, in particular interactions with the stellar wind, is that if such interactions can be observed, they can help characterise not only the planet but also the stellar wind of cool dwarf stars, which form the majority of planet hosts known nowadays.
 
\begin{acknowledgments}
I thank the SOC of the IAUS372 for the invitation to participate in this symposium. In particular, I thank the SOC chairs, Gianna Cauzzi  and Ali Tritschler, for their impeccable organisation of this meeting during very difficult pandemic times. AAV has received funding from the European Research Council (ERC) under the European Union's Horizon 2020 research and innovation programme (grant agreement No 817540, ASTROFLOW).
\end{acknowledgments}

\def\apj{{ApJ}}    
\def\nat{{Nature}}    
\def\jgr{{JGR}}    
\def\apjl{{ApJ Letters}}  
\def\apjs{{ApJ Supp.}}    
\def\aap{{A\&A}}   
\def\mnras{{MNRAS}}
\def\aj{{AJ}}
\def\grl{{Geophysical Research Letters}}
\def\araa{{Annual Review of Astronomy \& Astrophysics}}
\def\ssr{{Space Science Review}}
\let\mnrasl=\mnras


\end{document}